\documentclass[12pt,twoside,journal,draftclsnofoot,onecolumn]{IEEEtran}

\usepackage{amsmath,amsfonts,bm,amssymb,psfrag,ifthen,color,subfigure}
\usepackage{mathrsfs}
\usepackage[justification=centering,font=footnotesize]{caption}
\usepackage{todonotes}

\allowdisplaybreaks[4]

\usepackage[dvips]{epsfig}
\usepackage[dvips]{graphicx}
\usepackage{amsmath,amsfonts,bm,amssymb,psfrag,ifthen,color,subfigure}
\usepackage{algpseudocode}
\usepackage{algorithm}
\usepackage{algorithmicx}
\usepackage{ifthen}
\usepackage{setspace}
\usepackage{cite}
\usepackage{url}
\usepackage{longtable}
\usepackage{stfloats} 
\usepackage{graphics,booktabs,color,subfigure}
\usepackage{float}
\usepackage{diagbox}

\usepackage[dvips]{epsfig}
\usepackage[dvips]{graphicx}
\usepackage{amsmath,amsfonts,bm,amssymb,psfrag,ifthen,color,subfigure,amsthm}
\usepackage{algpseudocode}
\usepackage{algorithm}
\usepackage{algorithmicx}
\usepackage{setspace}
\usepackage{cite}
\usepackage{url}
\usepackage{longtable}
\usepackage{stfloats} 
\usepackage{graphics,booktabs,color,subfigure}
\usepackage{float}
\usepackage{diagbox}
\usepackage{multirow}
\usepackage{makecell}
\allowdisplaybreaks[4]

\newtheorem{Definition}{Definition}

	\usepackage{graphicx}

	\usepackage{graphicx}
	\usepackage{todonotes} 

\begin{document}
\title{Mathematical Characterization of Signal Semantics and Rethinking of the Mathematical Theory of Information}
  \author{Guangming Shi, \IEEEmembership{Fellow, IEEE}, Dahua Gao, Shuai Ma, Minxi Yang, Yong Xiao, and Xuemei Xie
   \thanks{Guangming  Shi and Dahua Gao are with
				Pengcheng Laboratory, Shenzhen, 518066, China and also
   with the				School of Artificial Intelligence, Xidian University, Xi'an, Shaanxi 710071,  China, (e-mail: gmshi@pcl.ac.cn,  dhgao@xidian.edu.cn).}
   \thanks{Shuai Ma is with PengCheng Laboratory, Shenzhen 518066, China. (e-mail:  mash01@pcl.ac.cn).}
      \thanks{ Minxi Yang and Xuemei Xie are with
the School of Artificial Intelligence, Xidian University, Xi'an 710071,
China. (e-mail:mxyang@stu.xidian.edu.cn, xmxie@mail.xidian.edu.cn).}

    \thanks{Yong Xiao is with the School of Electronic Information and Communications at
the Huazhong University of Science and Technology, Wuhan 430074, China,
also with the Peng Cheng Laboratory, Shenzhen, Guangdong 518055, China,
and also with the Pazhou Laboratory (Huangpu), Guangzhou, Guangdong
510555, China (e-mail: yongxiao@hust.edu.cn). }

}
\maketitle
\begin{abstract}

   Shannon  information theory is established based on probability and bits, and  the communication technology based on this theory realizes the  information age.    The original goal of Shannon's information theory is to describe and transmit information content. However, due to information is related to cognition, and cognition is considered to be subjective,   Shannon  information theory is to describe and transmit information-bearing signals.
    With the development of the information age to the intelligent age, the traditional signal-oriented processing needs to be upgraded to content-oriented processing.  For example,  chat generative pre-trained transformer (ChatGPT) has initially realized the content processing capability based on massive data.
      For many years, researchers have been searching for the answer to what the information content in the signal is, because only when the information content is mathematically and accurately described can information-based machines be truly intelligent.

  This paper starts from rethinking the essence of the  basic concepts of the information, such as semantics, meaning, information and knowledge, presents the mathematical characterization of the information content, investigate the relationship between them, studies the transformation from Shannon's signal information theory to semantic information theory, and therefore proposes a content-oriented semantic communication framework.
Furthermore,  we propose semantic decomposition and composition scheme to achieve conversion between complex and simple semantics.
 Finally, we verify the proposed characterization of information-related concepts by implementing evolvable knowledge-based semantic recognition.

\end{abstract}


\IEEEpeerreviewmaketitle
  \section{Introduction}


\subsection{Background}

 In 1948, Claude
Shannon used bits to describe messages, and probabilities to describe the information content of messages\cite{Shannon_1948}. The measurement method of messages and information content is the cornerstone of the development from industrialization to informatization. Looking back at the development history of communication technology, it is not difficult to find that the traditional communication mode mainly meets the business needs of human beings for obtaining information and enhancing communication experience\cite{Saad_INW_2020,Zhang_Engineering_2022}. Throughout the evolution of communication technology, starting from the initial telegraph systems to the incoming sixth generation (6G) technology, the communication business has undergone a significant transformation. This has resulted in a shift from basic text-based telegrams to advanced telephonic systems that are capable of comprehending tone and attitude of both parties during communication. This progression has further advanced to include visual communication mediums, including images that enable the clear visualization of expressions and actions of the other party. Additionally, the communication business has expanded to incorporate video calls, which facilitate real-time dynamic interaction and 3D virtual reality (VR) experiences. The latest advancement in communication technology includes the development of augmented reality (AR) services with comprehensive environmental perception \cite{Strinati_TVT_2019,Akyildiz_ITU_2022}. With the evolution of communication technology, the channel capacity and communication rate have been continuously improved, and the emotional experience of communicators has gradually been enriched. Communication technology serves a dual purpose in facilitating the transmission of large quantities of information between humans, as well as between humans and machines, and among machines themselves. Furthermore, communication technology has the capability of enabling immersive emotional interactions with a profound sense of presence\cite{Lueth_2020,Antoniou_2021}. The purpose of conventional communication technology is to accomplish precise transmission of messages and high-fidelity reproduction of signals. In this regard, the state of things conveyed in the messages and the emotional information contained therein are extracted and perceived by agents equipped with information processing capabilities.

In 1948, in his classic ``A mathematical theory of communication" \cite{Shannon_1948}, Shannon stated at the beginning that  the basic problem of communication is to exactly or approximately reproduce in one place a message selected in another place. These messages usually have specific meanings and are associated with specific physical or conceptual entities. The only thing that traditional communication technology cares about is how to accurately transmit the source signal to the destination without distortion, and does not care about what content is conveyed.

In 1949, Weaver, an information theory scholar who cooperated with Shannon, pointed out in the book ``Recent Contributions to the Mathematical Theory of Communication" \cite{Weaver_1949}: There are three levels of communication: the first is the signal level, the second is the semantic level, and the third is the information level. Scientists at the time believed that semantics did not have the universality required for communication, and the mathematical description of semantics was a very esoteric and complex issue. Since it is very difficult to extract the semantics of messages from signals by machines, and  people can extract semantics by themselves, traditional communication technologies do not involve semantic processing. Therefore, Shannon believes that in communication technology, we should avoid considering semantics, because semantics will greatly reduce the universality of communication theory. Shannon  information theory has always been the basic follow of communication technology, and it has guided the development of communication technology in recent decades with progress. It should be said that Shannon  information theory is one of the greatest scientific theories in the past 100 years. Under the guidance of Shannon  information theory that does not pay attention to the content of the message, we developed the purpose of increasing the amount of messages transmitted and enhancing the sense of experience at the cost of expanding bandwidth and increasing energy consumption.

 \subsection{Motivations}

As industrial informatization gradually drives the development of intelligence, human society has moved from the information age to the intelligence age, from the internet of everything to the intelligent connection of everything \cite{WANG_WC_2020}. Communication services have evolved from satisfying human-to-human communication to information interaction between human and machine and machine to machine. So is the need for machine-to-machine communication the same as that for everyone? As far as the current state of science and technology is concerned, the emotional mechanism of robots is not yet clear, and machines have no emotional needs. The main purpose of communication between them is to obtain information. Moreover, if a machine receives a message containing a large amount of useless information, it will not only waste communication resources, but also consume the computing power and time of the machine to process massive data, resulting in the inundation of valid information and making it impossible to make a quick decision. For example, for the operating robots on the production line, the lag of the optimal production scheduling decision may lead to low production efficiency, affecting product shipments, yield and quality indicators; another example, in the fierce confrontation battlefield, the weapon system Key information that cannot be quickly extracted from the wide area sensor network may lose the best time to strike or be destroyed by the enemy's preemptive strike.
   Therefore, the main purpose of communication is information delivery rather than message delivery\cite{Shi_CM_2021}. Since the signal carries the information required for decision-making, the machine needs to extract the information in the signal and understand the content of the information to eliminate the uncertainty of the information required to perform the task. However, apart from human voice and text, which can directly carry the information obtained by human processing the signal of things, most of the information transmitted by the current communication system is the information perceived by the sensor, not the information required for intelligent decision-making. So, can the communication system be directly oriented to the transmission of information? Up to now, there is still no set of theories that can describe and measure the content of information.

If the key information that can be used for decision-making is directly transmitted to the agent, the communication time will be greatly shortened under the same communication resource conditions. The communication terminal saves the process of extracting useful information from redundant messages, which is more conducive to real-time decision-making and rapid response of the communication terminal. Therefore, we have to rethink Shannon  information theory, whether this classic theory that has long guided and served human-to-human communication needs to be expanded and extended to guide the development of content-oriented communication technology in the intelligent age.

 Shannon introduced ``bits" as the unit of measurement for information and employed the probability of event occurrence to quantify the amount of information. It was noted that the amount of information is inversely correlated with the probability of event occurrence.
 In the intelligent era, Shannon information theory lacks reasonable explanations for many problems.
	\begin{itemize}

  \item \emph{Shannon information theory ignores signal complexity \cite{Cover_Book}:}  ~For an analog signal with complex dynamic characteristics, the bit stream is obtained  based on  Nyquist sampling  theorem. Generally, the sampling frequency of complex signals is  higher than that of simple signals, and thus the amount of bits after sampling is more. Because the amount of information defined by Shannon entropy is only related to the probability of event occurrence, it is independent of the complexity of the signal. Then, according to Shannon  information theory, the information content of a high-complexity signal and a low-complexity signal can be exactly the same, which is contrary to human intuition (complex signals should have more information than simple signals).

 \item \emph{ Shannon information theory cannot measure semantic information\cite{Bao_INSW_2011}:}~  Assume that in a paragraph of text, there is a common keyword, whose   content of the keyword is very important,   and while there is a rare modal particle, whose content is unimportant and  only helps to understand the connotation of the text effect. How to explain the information content of this unimportant modal particle is more than that of the important keyword?

 \item \emph{Shannon  information theory cannot explain why the same message has different semantic information in different contexts\cite{Weaver_1949,Lan_JSIN_2021}:}~
  For the news of the same content, people with different knowledge backgrounds and situations feel different amounts of information after receiving it. If the information sent surprises the recipient or is significantly different from the information the recipient has already mastered, the recipient thinks the amount of information is large, and vice versa. For message content that appears with the same probability, the information content is exactly the same, but the information content expressed can be completely different. Different information acts on the same receiver with different strengths and consequences, that is, the utility of the information is completely different. Shannon  information theory cannot explain why messages with the same amount of information can have completely different information utility?

\end{itemize}
The fundamental reason is that Shannon information theory does not involve the meaning of message, so it cannot explain the above problems \cite{Kountouris_CM_2021}.
Scientists in the Shannon era realized the role of semantics in communication, but they could not uniformly describe semantics mathematically, nor did they have algorithms and hardware devices that could extract semantics. Moreover, the scholars at that time believed that semantics was a subjective feeling in people's hearts. Everyone's definition and feelings of semantics were different, and they were not deterministic and unique\cite{Vatansever_NC_2021}. Therefore, semantics are not considered nor are semantics involved in communication systems. The essence of communication is to transmit content, and to grasp the content, you must first describe its semantics. Is semantics really only subjective and not unique? Researchers  explored the relationship between the human brain and semantics by detecting the cortical responses of subjects to the semantic content of texts \cite{Huth_2016}. They found that there is a ``semantic system" composed of a series of brain regions in the human brain to encode semantic information. The human brain has a semantic system for semantic perception and storage, which is widely distributed in the cortex of more than 100 different regions in the two hemispheres.   By studying the manifestation of semantics in the human brain, it can be seen that semantics is perceivable, learnable, and expressible by humans, and is the basic unit of human expression information. Semantics exists objectively, and its meaning can be standardized and unique after convention.

With the development of the information age to the intelligent age, the traditional signal-oriented processing needs to be upgraded to content-oriented processing.  For example,
   in November 2022, OpenAI launched a  new large language model  (LLM): chat generative pre-trained transformer (ChatGPT) \cite{ChatGPT}, which has initially realized the content processing capability based on massive data. Specifically,  based on an improved version of GPT-3  \cite{BROWN_2020},   ChatGPT   acquired 100
million monthly active users in January 2023 due to its capability for content-oriented processing,    such as
medical report simplification \cite{Katharina_2022},
bug fixing in computer programs \cite{Dominik_2023}, and stance detection from texts \cite{Bowen_2022}.
 In short, in the era of intelligence, in order to meet the efficient information exchange between agents, it is necessary to study new theories and new methods on how to transmit semantic information\cite{Xie_TSP_2021,Weng_JSAC_2021,Bourtsoulatze_TCCN_2019}.
So far,    the fundamental theory of semantic communications is still in its infancy with many open challenges\cite{Sana_CCNC_2022,Shi_CM_2021,Luo_WC_2022,WILLEMS_ISIT_2005}.
   Specifically, the definition of basic concepts of the semantics    has not yet been unified\cite{Dumais_2004,GIRSHICK_CVPR_2014,CHUTE_MBSV_1991,TURNEY_JAIR_2010,MATSUNO_BS_1992},  because the     mathematical definition of semantics is a very profound and complex problem.

\subsection{Contributions}

  With the above motivations,
 we  rethink  the essence of the basic concepts of the information, such as
semantics, meaning, information and knowledge, presents the mathematical characterization of the information
content, investigate the relationship between them, studies the transformation from Shannon's
signal information theory to semantic information theory, and   proposes a content-oriented
semantic communication framework.
  Specifically,
  the main contributions of our work are summarized  as follows:

	\begin{itemize}
 \item
According to the messages sources, we divide the messages into real-scene messages and virtual-scene messages for the first time, which   is to describe the subjective characteristics of the agent in an objective way.
Real-scene message is the signal obtained by sensors detecting objective things, which is the perception of the spatio-temporal state of objective things, while virtual-scene message is the signal expressed by human beings in language, text, and graphics, which is the subjective description of the spatio-temporal state of objective things by the human brain.

\item Furthermore, we propose definitions of semantics, meaning, information and knowledge for semantic communications, and characterize the relationship among them. Specifically,  meaning is the result of the brain's understanding of the spatio-temporal state of objective
things.
information is the image of the part of the message that the message receiver does
not know in advance and can decode and understand its meaning.
 Semantics is the label or naming of the characteristic function of the virtual-scene message to the characteristic function of the real-scene message, and knowledge  is the law of the spatio-temporal state of things discovered or created by human beings, which can be expressed by semantic symbols. Both semantics and knowledge have   hierarchical structures, where semantics is the unit of expressing knowledge, and knowledge is the result of semantics expressed hierarchically or in parallel.

 \item We propose semantic decomposition and composition to achieve conversion between complex and simple semantics. Specifically, we achieve semantic decomposition by solving a set of semantically simpler bases based on complex semantics. We propose an evaluation criterion for simple semantics as base functions from the perspective of reducing the complexity and error after decomposition. The semantic decomposition problem is formulated as a bi-level optimization problem based on this criterion. By specifying the coefficients of the semantic bases, it is possible to generate complex semantics based on simple semantics.

\end{itemize}

	The rest of this paper is organized as follows.
		Section II provides mathematical characterization of message, meaning, semantics, information, and knowledge.
		In Section III, we present the semantic decomposition and composition.
  		In Section IV, we experimentally validate the characterization of information-related concepts. Finally, we conclude the paper in Section V.

   \section{Mathematical characterization of message, meaning, semantics, information, and knowledge}

  Although
  there are many different definitions or explanations,    the  reasonable  definitions and mathematical  measurement  of   of basic concepts  of the  content-oriented processing
are yet   open problems\cite{Stratonovich,Howard,Ahlgren_CM_2012,Bisdikian_TSN_2013}.
In the intelligence era,   to meet the requirements of the efficient information exchange between agents, it is necessary to study new theories and  methods for characterizing signal content.   Therefore, we investigate
     mathematical characterization of basic concepts  of   content-oriented processing, such as message, information, semantics, knowledge\cite{Nieh_nature_2021},  meaning\cite{Nieh_nature_2021,Vatansever_NC_2021},  and   the relationship among them.

  \subsection{Mathematical characterization of message}

\emph{Message is spatio-temporal states of sensible or abstract things.} Spatio-temporal states are properties of things that vary with time and space, such as the temperature of a cup of coffee decreasing with time. Spatio-temporal states apply not only to sensible things, such as a cup of coffee, but also to abstract things, such as changes in the economy over time. Rules are a special class of abstract things, which tend to be spatio-temporally universal, i.e., their properties do not change with time or space, e.g., "1+1=2" always holds in any space-time. In this paper, we focus on sensible things and characterize a message as the set of their spatio-temporal states. Specifically, assume that the set of all states  is represented by the $\Omega$, and message, information, knowledge, and message unknown parts can be expressed as ${\rm{M}}$, ${\rm{I}}$, ${\rm{K}}$, ${\rm{D}}$, respectively. A message is a combination of these states, i.e,  ${\rm{M}} \subset \Omega $.

       \begin{figure}[htbp]
      \centering	
	\includegraphics[height=5cm]{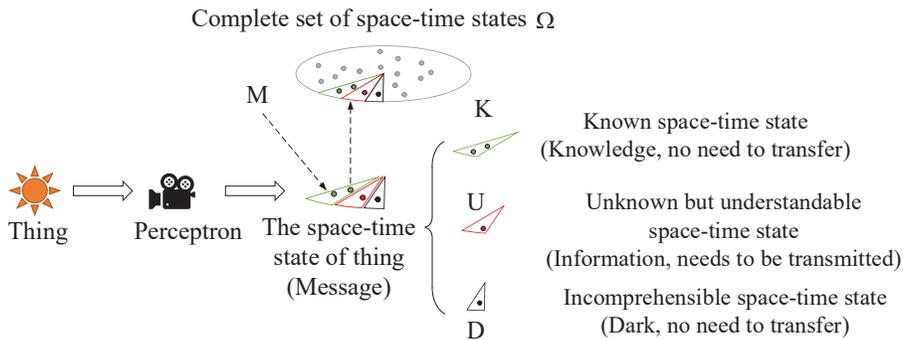}
   \caption{
   The  relationship between the spatio-temporal state of things and messages, semantics, information, and knowledge.}
   \label{state_knowledge}
\end{figure}

 Based on the above assumptions, we obtain the relationship   as follows
\begin{align}{\rm{M = I}} \cup {\rm{K}} \cup {\rm{D}} \subset \Omega.
\end{align}
The intersection of information characterization state and knowledge characterization state is an empty set, and the intersection with $D$ is also an empty set, i.e.,
\begin{align}{\rm{I}} \cap {\rm{K = }}\emptyset,~{\rm{I}} \cap {\rm{D = }}\emptyset .
\end{align}

Messages can also be represented by functions. Specifically, let $f\left( {X,t} \right)$ denote the state of things, where $t$ is the time dimension variable, and $X$ is the state variables of other dimensions.
     The message is a profile of the state of things, and a projection of $f\left( {X,t} \right)$.
     Furthermore,
    $f\left( {X,t} \right)$ can be    projected into multi-modal signals by sensors, such as
    sound wave signal ${\rm{Sound}}\left( t \right)$, light wave image signal ${\rm{Image}}\left( t \right)$, time light wave video signal ${\rm{Video}}\left( X, t \right)$, radar detection electromagnetic wave signal ${\rm{EleWave}}\left( X, t \right)$, tactile electrical signal ${\rm{Touch}}\left( X, t \right)$, and other multi-modal forms. Note that,
              ${\rm{Sound}}\left( t \right)$,  ${\rm{Image}}\left( t \right)$, ${\rm{Video}}\left( X, t \right)$, ${\rm{EleWave}}\left( X, t \right)$ and ${\rm{Touch}}\left( X, t \right)$ are all objective real signals.

        Moreover, if  a message was generated in the brain,   it describes the spatio-temporal state of subjective things.
The signals describing the spatio-temporal state $f\left( {X,t} \right)$ of things include voice ${\rm{Voice}}\left( t \right)$, writing  ${\rm{Writing}}\left( t \right)$ and painting ${\rm{Painting}}\left( t \right)$  signals, which are all signals controlled by the brain.
          Note that, the sound wave signal ${\rm{Sound}}\left( t \right)$ is different from the voice signal  ${\rm{Voice}}\left( t \right)$, where
           ${\rm{Sound}}\left( t \right)$ is the signal obtained by using sound waves to detect the state of things, and ${\rm{Voice}}\left( t \right)$ is the signal generated by the brain to describe the state of things with voice.
                     For example, the sound of ``bang" from beating a gong and the sound of people using language to describe the beating of a gong, where the former is a sound wave signal, and the latter is a voice signal.

Due to the limited perception ability of sensors,  a message generally can only describe a part of the state of things. For example, video signal ${\rm{Video}}\left( X, t \right)$ contains two modal signals of sound ${\rm{Sound}}\left( t \right)$ and image ${\rm{Image}}\left( t \right)$.
Generally, the spatio-temporal state of things can be projected into a multi-modal signal function by multiple types of sensors, and the signal $S\left( {X,t} \right)$ has the following expression
\begin{align}S\left( {X,t} \right) = \left( {Signa{l_1}\left( {X,t} \right),Signa{l_2}\left( {X,t} \right),...,Signa{l_n}\left( {X,t} \right)} \right),\end{align}
where the signal function $S\left( {X,t} \right)$ is the mathematical expression of the message.

The message is used to describe the spatio-temporal state of things, or the content of the message is the spatio-temporal state of things. A message is a collection of spatio-temporal states of multiple things.
It is well known that  information comes from messages.  The message is independent of the perceiver, while the information depends on the perceiver, and the message becomes information only after being perceived.

There are two kinds of things in the world, one is objective things, which exist in nature; the other is subjective things, which are constructed from the human brain. Therefore, there are two kinds of messages in the world, {real-scene messages  }and  {virtual-scene messages }, where   real-scene messages are description of the spatio-temporal state of objective things, while virtual-scene messages  are description of the spatio-temporal state of subjective things.

\subsubsection{Real-scene message and its content} The content of the {real-scene message} is the characteristic signal of the message. The content of the message represents the spatio-temporal state of things, and   what reflects the spatio-temporal state of things is its characteristic signal, which means that the content in the message corresponds to the characteristic signal one-to-one.

\begin{Definition}
 {Real-scene messages} is the signal obtained by actively or passively detecting objective things with sensors, and human beings feel the sensory signals to gain an understanding of the spatio-temporal state of things.
 \end{Definition}
 For example, we experience the world through light wave signals, sound wave signals, electromagnetic wave signals, smell, and touch signals, etc. The   content  of a    real-scene message   includes both local details and overall. For example, a photo of a lakeside contains not only macroscopic content, such as the outline of the lake and vegetation by the lake, but also microscopic content, such as waves on the lake surface and types of vegetation.
The content of real-scene news depends on the cognitive granularity of the perceiver. The content levels of different granularity cognition are different, and the corresponding  messages characteristic functions      are also different.
Note that, if the message is a real-scene message,   $S\left( {X,t} \right)$ is the multi-modal projection of the spatio-temporal state function $f\left( {X,t} \right)$ of objective things.

   \subsubsection{Virtual-scene message and its content}Similar to real-scene message, the characteristic signal of the spatio-temporal state image of things describes the content of virtual-scene messages. The image feature signal corresponds to the feature signal of the real scene, but it cannot be observed. Human beings use language, text, body, gesture, expression and other signals that can be output by themselves to describe the  spatio-temporal state image of things formed in the brain, that is, to use language, text, body, gesture, expression and other signals to name the image characteristic signal, which   is also the name of the real-scene characteristic signal.

   \begin{Definition} {Virtual-scene message} is the message expressed and transmitted by human beings in language, text and graphics. \end{Definition}

     It is the image of the spatio-temporal state of objective things in the human brain, or the spatio-temporal state of things imagined in the brain. For example, when a teacher is giving a lecture in the classroom, the language sound waves heard by the students are virtual-scene messages.
 There are two sources of virtual-scene messages: one is the state that the sender   understands after perceiving the spatio-temporal state of objective things; the other is the spatio-temporal state of things subjectively fabricated by the sender. It can be said that real-scene message is a ``thing", while virtual-scene message is an ``image".
Note that, the characteristic function of the real-scene messages is related to the purpose, not fixed.
  Specifically, different angles of observing things have different characteristic functions of real-scene messages.
For example, the characteristic function of the signal can be the mean value, the derivative function, or the variance function and so on.
The characteristic function of the virtual-scene messages is fixed, because the virtual-scene information is the original information processed by the brain to form a fixed or agreed state, such as language, gestures and text.


 If the message is a virtual-scene message, then $S\left( {X,t} \right)$ describes the state of subjective things.
Since the subjective thing is the image of the objective thing in the human brain, in order to distinguish it from the  spatio-temporal state of the objective thing, we use the virtual image function  $f'\left( {X,t} \right)$ to represent the  spatio-temporal state of the subjective thing. The virtual image function  $f'\left( {X,t} \right)$ is constructed by the human brain, which is the mirror image of  $f\left( {X,t} \right)$ in the brain.

     In this paper, for the convenience of description,   we also   the space-time state function $f\left( {X,t} \right)$ of objective things with the real-scene message signal $S\left( {X,t} \right)$. In practice, it is impossible for us to obtain the full $f\left( {X,t} \right)$, because the sensor's perception capacity is always limited.

        \begin{figure}[htbp]
      \centering	
	\includegraphics[height=5cm]{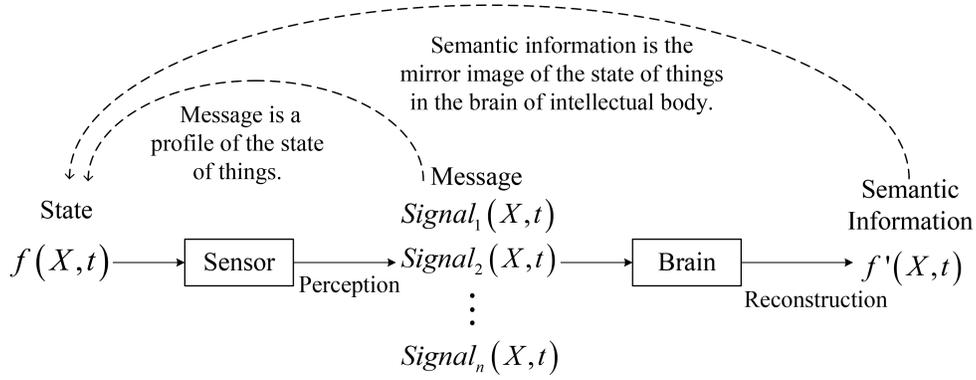}
   \caption{
   The  relationship between the state of things of the real-scene and the state of things of the virtual-scene.}
   \label{real_and_virtual_states}
\end{figure}

\subsection{Mathematical characterization of meaning}

Meaning is the result of the brain's understanding of the  spatio-temporal state of objective things, that is, the virtual image function   $f'\left( {X,t} \right)$. Figure \ref{real_and_virtual_states} demonstrates this process, where   $f'\left( {X,t} \right)$ approaches to  $f\left( {X,t} \right)$, and they could be different.
 The so-called understanding is the process by which the brain perceives the meaning of the message, and the corresponding mathematical characterization of understanding is given as
  	\begin{itemize}
 \item If $f'\left( {X,t} \right) = f\left( {X,t} \right)$, it is a correct understanding;
\item If $f'\left( {X,t} \right)  \ne  f\left( {X,t} \right)$ it is a misunderstanding.
  	\end{itemize}

\subsection{Mathematical characterization of semantics}

There are   many definitions and explanations on semantics. So far,   in computer vision, communication, or other information fields, scholars   generally used the  ``semantic" word     to describe the meaning of information carried in signals, such as semantic segmentation \cite{GIRSHICK_CVPR_2014}, semantic analysis \cite{CHUTE_MBSV_1991}, semantic understanding \cite{TURNEY_JAIR_2010}, or semantic computation \cite{MATSUNO_BS_1992}. However, what is semantics in the field of information, and how to explain semantics? How to characterize and measure semantics? What is the nature of semantics? And so on, the above questions are not fully answered.

  In this paper, we  define  semantics based on the analysis of real-scene and virtual-scene messages. Intuitively, semantics is the result of the brain's perception of a real-scene signal describing the state of objective things. In essence, it is an image formed by the brain to describe a spatio-temporal state of things on the signal. This image needs to be expressed through virtual-scene information and virtual-scene signals. The characteristic signal in the real-scene signal is denoted as ${f_{{\rm{real}}}}$ to describe the spatio-temporal state of objective things, and the characteristic signal of the virtual-scene signal is denoted as ${f_{{\rm{virtual}}}}$ to express the image of the corresponding state. ${f_{{\rm{virtual}}}}$ may be a characteristic signal of speech modality, and it may also be a characteristic signal of text, graphics and other human senses that can express modality.

The  semantics is the label or naming of the virtual-scene message characteristic function to the real-scene state message characteristic function.
 \begin{Definition}
{Semantics} is the characteristic function ${f_{{\rm{real}}}}$ of the spatio-temporal state of things in real-scene messages, which is named after the characteristic function ${f_{{\rm{virtual}}}}$ of the modal signal of virtual-scene messages.
  \end{Definition}
Similar to data that can be labeled, the spatio-temporal state function ${f_{{\rm{real}}}}$ of the thing has a name ${f_{{\rm{virtual}}}}$, and ${f_{{\rm{virtual}}}}$ has an attribute ${f_{{\rm{real}}}}$. This naming and attribute relationship can also be expressed with semantic notation. Semantic symbols are the bridge between virtual-scene messages and real-scene messages.

Note that, in general, semantics is ambiguous. The semantics we study refers to the knowledge framework of a given domain, so as to ensure that there is no singularity in the semantics. An ${f_{{\rm{real}}}}$ can have multiple ${f_{{\rm{virtual}}}}$ names, but one ${f_{{\rm{virtual}}}}$ name only corresponds to one ${f_{{\rm{real}}}}$. In reality, there is a situation where one f-fiction corresponds to multiple ${f_{{\rm{real}}}}$. Words are polysemy because they span different fields. For example, there are many roads named ``Zhongshan Road" in cities all over the country, but there will not be two ``Zhongshan Roads" in one city. Thus, we can find that the essence of semantics is signal features, but signal features are not necessarily semantics.

Assume that the characteristic function of the  spatio-temporal state describing the real-scene is ${F_i}\left( {X,t} \right)$, and the modal signal expressing this spatio-temporal state in the brain, such as a voice signal, has a characteristic function of ${P_i}\left( {X,t} \right)$. By using  ${P_i}\left( {X,t} \right)$ to name ${F_i}\left( {X,t} \right)$, and ${F_i}\left( {X,t} \right)$ to  ${P_i}\left( {X,t} \right)$ to explain attributes, ${F_i}\left( {X,t} \right)$ and  ${P_i}\left( {X,t} \right)$ establish a naming-attribute similar ``semantic association pair", which is semantics, and can be represented by semantic symbols: $\$$ as follows:
\begin{align}{\$ _i} = \left\{ {{F_i}\left( {X,t} \right),{P_i}\left( {X,t} \right),\ldots} \right\},
 \end{align}
where   ``$...$" indicates that there may be other modal characteristic signals associated together.

\begin{figure}[htbp]
      \centering	
	\includegraphics[height=5cm]{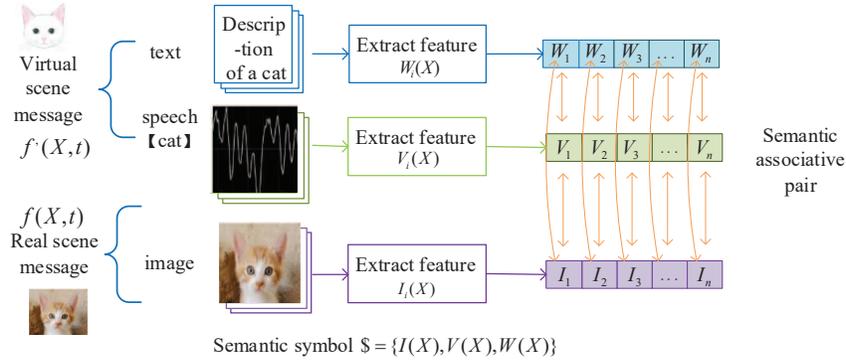}
   \caption{
   The  relationship between the state of things   and meaning.}
   \label{real_virtual_pair}
\end{figure}

Semantics is a characteristic function of virtual-scene message with properties represented by characteristic functions of real-scene messages, as shown in Fig. \ref{real_virtual_pair}. Semantics can be associated from a characteristic function of one modality to a characteristic   function of another modality, which is the  spatio-temporal state  of real-scene things.  Hence, semantics is also the expression of the  spatio-temporal state of things, which can be  understood by a group of people.

Meaning is the spatio-temporal state of subjective and objective things reconstructed by the brain using semantic symbols and their structural relationships for messages, i.e., $f'\left( {X,t} \right)$.

\subsection{Definition   of Information}

Information comes from the message, and the information is the    perception result of the message.
 \begin{Definition} {Information} is the image of the part of the message that the   receiver does not know in advance and can decode and understand its meaning.  \end{Definition}
 If the  receiver knows all the content of the message before receiving the message, the received message      has no information, and knowledge can be considered as the image of a known message. To be able to decode means that information should be   encoded with semantic symbols known to the receiver, and enable the receiver to understand the meaning. There is also a part of the message that the receiver cannot decode with known semantic symbols, which is called \emph{{dark  message}}.

\subsection{Mathematical Characterization of Knowledge}

 \begin{Definition}
{Knowledge} is the law of the spatio-temporal state of things discovered or created by human beings, which can be expressed by semantic symbols, or by characteristic functions ${f_{{\rm{real}}}}$ and ${f_{{\rm{virtual}}}}$.
 \end{Definition}

 The knowledge expressed by semantic symbols or ${f_{{\rm{virtual}}}}$ is the well-known knowledge graph, and the knowledge represented by ${f_{{\rm{real}}}}$  is called library function. Knowledge base functions can be  used for knowledge  calculations, while knowledge graphs can only be used for retrieval.
Both semantics and knowledge have a hierarchical structure, where semantics is the unit of expressing knowledge, and knowledge is the result of grouping and expressing many semantics according to levels and juxtapositions.  Both  knowledge map and  knowledge function library are   hierarchical structures, that is, high-level knowledge is combined or calculated from the underlying knowledge.


  \subsection{The relationship between     message, semantics, meaning information and knowledge.}


\begin{table*}[ht]
 \caption{The relationship between message, semantic, information, knowledge and meaning.}
 \label{Technical comparison}
 \centering
 \begin{tabular}{|c|c|c|c|c|c|c|c|c|c|}
  \hline
  & \footnotesize  \makecell[c]{Real\\ view}&\footnotesize  \makecell[c] {Virtual \\ view} & \footnotesize  \makecell[c]{Known \\ in advance} & \footnotesize  \makecell[c] {Currently\\ known}& \footnotesize              \makecell[c]{Unknown \\beforehand}&\footnotesize  \makecell[c] {Presently \\unknown} &\footnotesize   \makecell[c]{Components} &\footnotesize  \makecell[c] { Overall} \\ \hline
    \footnotesize  Message& \footnotesize  \footnotesize \checkmark&  \footnotesize \checkmark&  & & & &  &  \footnotesize \checkmark \\ \hline
    \footnotesize  Dark messages&  \footnotesize \checkmark&  \footnotesize \checkmark&  & & &  \footnotesize \checkmark&  &  \\ \hline
    \footnotesize Semantic&  &  \footnotesize \checkmark&  \footnotesize \checkmark&  \footnotesize \checkmark&  & &  \footnotesize \checkmark&\\ \hline
    \footnotesize  Information&  &  \footnotesize \checkmark& &  \footnotesize \checkmark&  \footnotesize \checkmark & & &  \footnotesize \checkmark\\ \hline
    \footnotesize Knowledge&  &  \footnotesize \checkmark&  \footnotesize \checkmark& &  & &  &   \footnotesize \checkmark \\ \hline
    \footnotesize  Meaning&  &  \footnotesize \checkmark&  \footnotesize \checkmark&  \footnotesize \checkmark& \footnotesize \checkmark  & & &  \footnotesize \checkmark\\ \hline

 \end{tabular}
\end{table*}

Table I summarizes the relationship between message, semantics, information, knowledge, and meaning. Specifically,
semantics are the fundamental building blocks for expressing information, knowledge, and meaning. Semantics, information, knowledge, and meanings are all formed by the description of the state of things in the brain. Semantics is the image component that can understand the spatio-temporal state of things known in advance after receiving the message. Information is the image that can be understood immediately after receiving the message, and the corresponding  spatio-temporal state of things not known in advance. Knowledge is the image of the spatio-temporal state of things that can be understood but known before receiving the messages. Meaning is the image of the spatio-temporal state of things that can be understood immediately after receiving the messages, which may include the image of the spatio-temporal state of things known and unknown in advance.

         \begin{figure}[htbp]
      \centering	
	\includegraphics[height=5cm]{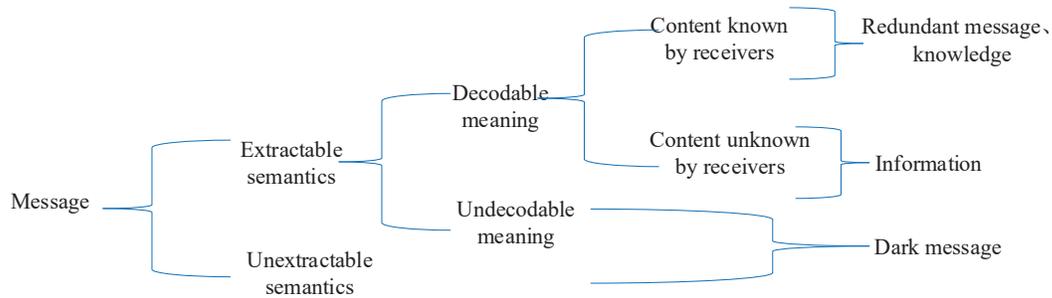}
   \caption{
   The classification relationship of message, semantics, meaning  and knowledge.}
   \label{classification}
\end{figure}

Fig. \ref{classification} shows the   relationship of message, semantics, meaning  and knowledge.
Specifically, a message may contain semantics, information and knowledge that the receiver can perceive, as well as content that cannot be perceived. For example, Alice transmits a message to Bob through voice, part of which is understood by Bob but known in advance, part of which is understood by Bob but unknown in advance, and part of which Bob has heard but does not understand.
The following example contains three levels as follow:
 	\begin{itemize}
 \item  \emph{Signal level:}  Bob can hear the voice from Alice;

 \item  \emph{Semantic level:} Bob is able to extract semantics from the signal, which is the expressive level of communication between Alice and Bob. If Alice talks to Bob in English, and Bob doesn't understand English, Bob won't be able to extract any semantics from Alice's speech;

    \item \emph{Information level:} The premise that Bob can expressly receive Alice's message is that Alice adopts Bob's information encoding symbols, which means that Bob understands every word Alice said.
          There are two scenarios in the information level as follows:\\
\emph{Scenario 1:}   Bob understands every word and the combined content; at this time, Bob can distinguish which content of the signal is known and which content is unknown in advance, that is, information.\\
\emph{ Scenario 2:}      Bob understands every word, but he does not understand the combined content. For example, Bob is a primary school student, and Alice teaches Newtonian mechanics to Bob, but Bob cannot decode it. This scenario has the same effect as Bob's inability to extract the semantic scenario from the signal, that is, neither can obtain information. The part of information that cannot be obtained is called dark message.

\end{itemize}
Therefore, knowledge corresponds to the known part of the message, and information is the content of the message that the receiver does not know in advance and can understand the meaning of. The information is absorbed by the recipient and transformed into new knowledge for subsequent communication. Information is the knowledge difference between before and after reception, and this difference is used to eliminate the uncertainty of the receiver's understanding of the original unknown.

Furthermore, the communications   comprise four
  levels as follows:
	\begin{itemize}
 \item  \textbf{Signal level}: The signal level focuses on accurately
transmitting bits over a noisy communication channel, which can be viewed as   syntax   communications.

 \item  \textbf{Semantic level}: The meaning level focuses on   precisely
transmitting semantic contents to the receiver,  which can be viewed as   semantic   communications.

 \item  \textbf{Informative  level}: The   informative  level focus on   precisely
transmitting unknown information  to the receiver, which can be viewed as informative  communications;

 \item  \textbf{Effectiveness level}: The effectiveness level focuses on how  effectively
 the received meaning affects    the control and decision-making of the receiver in the desired
way,    which can be viewed as intelligent pragmatic communication.

\end{itemize}

In summary, real-scene message is practical and objective, and a message is a set of characteristic functions describing the spatio-temporal state of multiple things. Virtual-scene information is subjectively sent from the brain, and its origin is the image of the real scene or a fictional image. Semantics, information, knowledge, and meaning are all reflected in the brain, and they all belong to the ``image" category. They are a set of various modal characteristic functions that express the spatio-temporal state of the virtual scene.

\section{Semantic decomposition and composition}
The complexity of the semantics of describing different things varies, with complex semantics containing more elements and features and requiring more interpretation and definition. For example, the semantic meaning of ``car" is more complex than the semantic meaning of ``tire". Complex semantics are composed of simple semantics. For example, ``car" consists of ``tires", ``engine", ``windows" and other parts. Breaking them down into simple semantic components can reduce the complexity of understanding and communication. It also allows for a deeper understanding of the composition and characteristics of things, so that new complex concepts and ideas can be created by recombining simple semantic components. For example, combining the semantic meanings of ``car" and ``bedroom" can create the new concept of ``caravan". In this section, we will introduce semantic decomposition and composition.

\emph{  The semantic operations, such as decomposition and composition,  are only used for the calculation of the characteristic function of the real-scene message, while the characteristic function of the virtual-scene message is only used to express the semantic symbol.
}


\subsection{Semantic decomposition}
In the previous section, we introduced semantics as a mutual labeling relationship between the real-scene and virtual-scene semantic characteristic functions. Since semantic characteristic functions are continuous functions that describe the spatio-temporal state of things. Then a complex semantics can be expressed as a combination of multiple simple semantics with residues:
\begin{equation}  \label{semantic_decomposition}
  F(X,t)=\sum_{n}{c_n f_n(X,t)+\sigma},
\end{equation}
where $F(X,t) $ and $f_n(X,t) $ are the complex and simple semantic characteristic functions, respectively, $c_n$ denotes the semantic decomposition coefficients, and $\sigma$ is the residual term. When $\sigma = 0$, the complex semantic characteristic function can be completely reconstructed using the simple semantic characteristic function and the semantic decomposition coefficients, which we call lossless semantic decomposition; conversely, when $\sigma \ne 0$ is lossy semantic decomposition. Simple semantic characteristic functions can be further decomposed until they are decomposed into semantic primitives. For example, matter can be decomposed into protons, neutrons, electrons and even quarks.

There may be different ways to decompose things from different perspectives of cognition. For example, there are Fourier series expansions, Taylor series expansions, and so on, for generic expansions of functions. Different decomposition methods use different basis functions. It can be said that the basis function determines the direction of decomposition, which also determines the perspective of cognitive things. Fourier series and other expansions are cognitive from a mathematical point of view, in which the basis functions have no semantics. To decompose complex semantics into simple semantics rather than mathematical series, one needs to solve the basis functions with semantic meaning, i.e., semantic bases.

\subsection{Criterion for semantic bases}
First, an evaluation criterion needs to be designed to guide the optimal solution of the semantic bases. We denote the probability distribution of a semantic characteristic function by $X$, a semantic characteristic function by $x$, satisfying $x\in X$, and the set of semantic bases obtained by decomposition is denoted as $K=\{k_i\}$. We suggest that the semantic base set can be evaluated in terms of the storage volume $V$, the average representation complexity $L$, and the average decomposition error $\mathcal{E}$. The storage volume is the size of the capacity required to store all semantic bases in the set. It represents the complexity of the set of semantic bases. The average representation complexity is the average size of the capacity required to store the semantic decomposition coefficients corresponding to the decomposition of all semantic eigenfunctions. It represents the complexity of describing the semantic characteristic function using this semantic basis set. The average decomposition error is the average error between the semantic characteristic function described using the semantic bases and the true value. It represents the accuracy of the set of semantic bases. The semantic base storage and average representation complexity should be as small as possible while satisfying the average decomposition error constraint, as:
\begin{equation}  \label{semantic_bases_criterion}
  \begin{split}
    & \underset{K}{\min}V(K)+\lambda L(K) \\
    & \text{s.t.}\ \mathcal{E}(K)<\epsilon,
  \end{split}
\end{equation}
where $\epsilon$ is the maximum tolerable average decomposition error and $\lambda$ is the trade-off weight between the storage volume and the average representation complexity.

Since complex semantics needs to be decomposed into semantic bases representing simple semantics, we assume that the distribution of complex semantics is a joint distribution composed of relatively independent semantic bases, as:
\begin{equation}  \label{joint_distribution}
  X=(X_1,X_2,\cdots,X_i,\cdots,X_J),
\end{equation}
where $X$ is the complex semantic distribution, $X_i$ is the semantic base distribution, and $J$ is the number of semantic bases.

The key problem of semantic decomposition is how to find the semantic bases that can represent different sub-distributions by complex semantic samples. We set the complex semantic decomposition into $N_k$ semantic bases $K=\{k_i\}_{i=1}^{N_k}$ based on a set of sample sets $D=\{x_i\}_{i=1}^{N_k}$. The semantic decomposition is to obtain a set of semantic bases with the smallest possible distributional correlation to represent the different sub-distributions under the condition that the cognitive error limit $\epsilon$ is satisfied. The corresponding optimization problem is as follows:
\begin{equation}  \label{inner_optimization_problem}
  \begin{split}
    & \underset{K}{\max}\mathcal{D}(K|D) \\
    & \text{s.t.}\ \frac{1}{N}\sum_{i=1}^N{\mathcal{E}(x_i|K)\le\epsilon},
  \end{split}
\end{equation}
where $\mathcal{D}(K|D)$ is the distribution diversity of the set of semantic bases $K$ given sample set $D=\{{{x}_{i}}\}_{i=1}^{N}$, expressed by the distribution distance between two pairs of elements in $K$:
\begin{equation}  \label{distribution_divergence}
    \mathcal{D}(K|D)=\frac{1}{{{N}_{K}}({{N}_{K}}-1)}\sum_{i=1}^{{{N}_{K}}}{\sum_{j=i+1}^{{N}_{K}}{d(k_i,k_j|D)}},
\end{equation}
where $d(k_i,k_j|D)$ is the distribution distance between ${{k}_{i}}$ and ${{k}_{j}}$:
\begin{equation}  \label{distribution_distance}
  d(k_i,k_j|D)=W(P({{k}_{i}},K),P({{k}_{j}},K)),
\end{equation}
where $W(\cdot ,\cdot )$ is a distance metric such as the Wasserstein distance, $P({{k}_{i}},K)\triangleq (P({{k}_{i}},{{k}_{1}}),\allowbreak \cdots ,\allowbreak P({{k}_{i}},{{k}_{j}}),\allowbreak \cdots ,\allowbreak P({{{k}_{i}},{{k}_{{N}_{k}}}}))$ defines the structural information of the semantic base ${{k}_{i}}$ in the full set $K$, $P({{k}_{i}},{{k}_{j}})$ is the joint probability distribution of ${{k}_{i}}$ and ${{k}_{j}}$ obtained by statistical estimation based on sample set $D$. $\mathcal{E}({{x}_{i}}|K)$ is the cognitive error of the signal ${{x}_{i}}$ according to the set $K$ of semantic bases:
\begin{equation}
    \mathcal{E}({{x}_{i}}|K)=\underset{{{c}_{1}},{{c}_{2}},\cdots ,{{c}_{{{N}_{K}}}}}{{\min }}\,{{f}_{e}}({{x}_{i}},\underset{j=1}{\overset{{{N}_{K}}}{\sum }}\,{{c}_{j}}{{k}_{j}}),
  \label{coginition_error}
\end{equation}
where ${{c}_{j}}$ is the coefficient corresponding to ${{k}_{j}}$. Minimizing the storage and the average representation complexity in the solution set of \eqref{inner_optimization_problem} will yield the optimal result for the semantic decomposition as:
\begin{equation}\label{sematic_decomposition_problem}
  \begin{split}
    \underset{K}{{\min }} &  \ v(K)+\lambda \frac{1}{N}\sum_{i=1}^{N}{l({{x}_{i}}|K)} \\
  \text{s.t.} & \  K\in \arg \underset{K}{\max}\{\mathcal{D}(K|D):\frac{1}{N}\sum_{i=1}^{N}{\mathcal{E}({{x}_{i}}|K)}\le \epsilon \},
  \end{split}
\end{equation}
where $v(K)$ is the storage volume of $K$, $l({{x}_{i}}|K)$ is the representation complexity $({{c}_{1}},{{c}_{2}},\cdots ,{{c}_{{N}_{K}}})$ based on $K$ for ${{x}_{i}}$, and $\lambda$ is the balance weight of the storage volume and the average representation complexity.

\subsection{Semantic hierarchy}
It is well known that human semantics is hierarchical, and simple semantics obtained from complex semantic decomposition can be further decomposed until they are decomposed into semantic primitives. For this reason, we propose a semantic decomposition that can be performed iteratively. For a first-order semantic base $k_{i}^{1}$, the set of samples $D_{i}^{1}$ that match only $k_{i}^{1}$ (i.e., $k_{i}^{1}$ corresponds to a decomposition coefficient $c_{i}^{1}$ much larger than the other decomposition coefficients) can be sampled from the environment. For example, for ``tires" of a ``car", many photos of only tires can be taken as a sample set. $k_{i}^{1}$ can be decomposed by solving \eqref{sematic_decomposition_problem} to obtain the set of second-order semantic bases $K_{i}^{2}=\{k_{i,j}^{2}\}_{j=1}^{N_{K}^{2, i}}$. For example, ``tire" can be further broken down into ``rubber" and ``wheel". In this case, the components of $k_{i}^{1}$ need to be replaced by the set of second-order semantic bases $K_{i}^{2}=\{k_{i,j}^{2}\}_{j=1}^{N_{K}^{2,i}}$ when calculating the storage volume and the average representation complexity. For example, the result of the decomposition of ``car" is ``window", ``engine", ``rubber", ``wheel", etc., where ``tire" has been replaced by ``rubber" and ``wheel". A decomposition can be called meaningful if it satisfies that the storage and average representation complexity metrics after decomposition are better than those before decomposition. Similarly, a second-order semantic base can be further decomposed into several third-order semantic bases and iteratively decomposed until the last meaningful decomposition is completed. For example, a ``tire" can be decomposed all the way down to ``rubber molecule" and ``air molecule". Because of the large number of molecules and the complex relationships between them, using microscopic particles to describe macroscopic things does not simplify the description, but significantly increases the complexity of the description. Therefore, this is not a meaningful decomposition. The optimized semantic hierarchy is obtained after all semantic bases are meaningfully and iteratively decomposed.

\subsection{Cognition based on semantic hierarchy}
Humans often categorize and compare new things based on existing knowledge to achieve knowledge of things. The optimized semantic hierarchy obtained after semantic decomposition can be considered as a form of knowledge. The process of categorization and comparison based on semantic hierarchy can be represented by the following optimization problem:
\begin{equation}  \label{semantic_cognition}
  c_1^*, c_2^*, \cdots, c_{N_K}^*=\arg \min _{c_1, c_2, \cdots, c_{N_K}} f_e(x_i, \sum_{j=1}^{N_K} c_j k_j),
\end{equation}
where $c_1^*, c_2^*, \cdots, c_{N_K}^*$ is a set of semantic decomposition coefficients that makes the cognitive error minimal. By recording or transmitting this set of coefficients instead of complex semantics, more efficient semantic understanding and semantic communication can be achieved.

\subsection{Semantic composition}
The typical intelligent behavior of humans is imagination. Humans create by combining things in the existing environment to form things that do not exist in the existing environment. New knowledge is discovered by experimenting to verify whether the created things are reasonable or not. Based on the knowledge, the existing cognitive system is modified. This is exactly the approach of human scientific research. Similarly, the hierarchical semantic base system constructed by using semantic decomposition can generate new semantics, and we call the process semantic composition. According to \eqref{semantic_decomposition}, specifying a different set of semantic decomposition coefficients $\{\hat{c}_n\}$ results in a complex semantics $\hat{F}(X,t)$ that is different from $F(X,t)$. For example, by increasing the size and hardness of the wheels in a car, off-road vehicles with greater off-road performance can be created. The new semantics $\hat{F}(X,t)$, verified by the environment, is understandable and not known in the past for the intelligences, i.e., knowledge. We call this process of generating knowledge by the intelligence itself without an external source of information as semantic composition-based knowledge discovery. For example, a newly designed off-road vehicle is tested in a field environment and the off-road performance is indeed improved. From the perspective of the intelligent community, intelligences communicate with each other and do not generate new knowledge, but only achieve knowledge spreading, while knowledge discovery based on semantic composition can discover new knowledge. Thus, it seems that semantic decomposition and synthesis may be beneficial to achieve a self-evolving intelligent population in human-like societies. We will explore knowledge discovery based on semantic synthesis based on multi-agent reinforcement learning in our future work.

  \section{Experiments  and discussions}

 \subsection{Semantic Recognition Based on Knowledge Evolution}
 Since 2012 when AlexNet \cite{krizhevsky2017imagenet}, a convolutional neural network (CNN) \cite{lecun1989backpropagation} model using GPU acceleration, first won the ILSVRC competition based on the ImageNet \cite{deng2009imagenet} dataset, mainstream image recognition methods have relied on end-to-end training of deep networks \cite{lecun2015deep}. However, deep networks are also generally considered to have three drawbacks: 1) poor interpretability due to black boxes, 2) high training data requirements due to many optimization parameters, and 3) poor generalization due to the tendency to fall into local optimal solutions. At present, it seems that it is difficult to achieve the human level of semantic recognition with strong interpretability, low sample demand and high generalization ability by end-to-end training of deep networks only. In the previous section, we discussed the importance of semantics and knowledge for intelligences. In fact, the use of knowledge-guided semantic feature extraction is the key for an intelligent body to be able to achieve target recognition with strong interpretability, low sample demand and high generalization capability. Specifically, the interpretability of the recognition process is ensured by extracting interpretable semantic features, the training sample size is reduced by introducing prior knowledge to guide semantic feature extraction instead of training from scratch, and the generalization ability is improved by avoiding model over fitting based on cross-task knowledge.

\begin{figure}[htbp]
  \centering	
  \includegraphics[width=15cm]{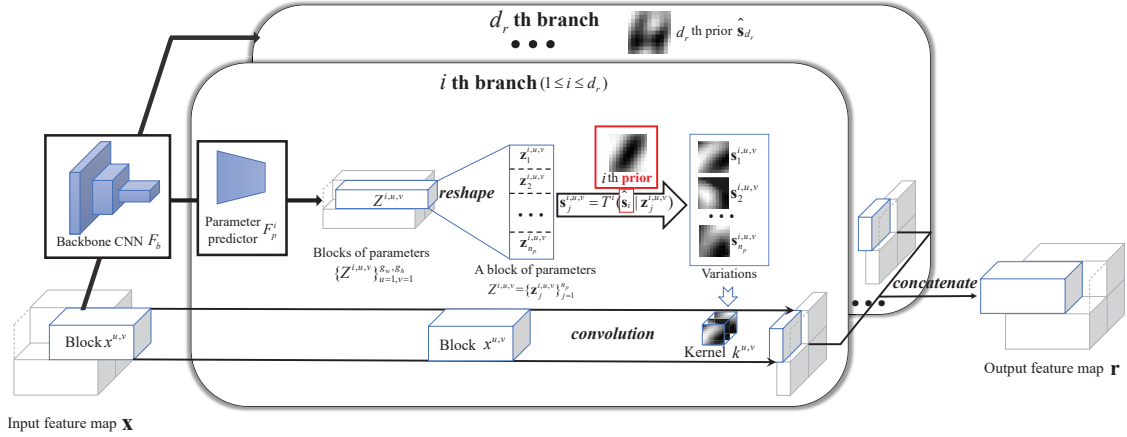}
  \caption{
    Structure diagram of MSConv, a semantic feature extraction module incorporating prior knowledge.}
  \label{fig_msconv}
\end{figure}

 We use the image semantic feature extraction module MSConv for fusing prior knowledge introduced in \cite{shi2022novel}, as shown in Fig \ref{fig_msconv}. MSConv requires a number of image patches representing prior knowledge, referred to as priors, as marked in the red box in the figure. First, the a priori corresponding parameters are predicted from the input feature map by a generic backbone CNN and parameter predictor based on the given priors. Finally, the output feature map is obtained by computing the convolution of the variations with the input feature map. In this case, the parameters required for the affine transformation are predicted from the input feature map, and each of its dimensions has a physical meaning, ensuring the interpretability of the feature extraction. As introduced in \cite{shi2022novel}, a simple and effective image classification network can be formed by sequentially connecting $1\times1$ convolution, global average pooling layer and one linear layer after for MSConv, which can achieve semantic recognition by fusing prior knowledge. In this experiment, we adopt the MSConv structure and network structure design consistent with \cite{shi2022novel}.

\begin{figure}[htbp]
  \centering	
  \includegraphics[width=8cm]{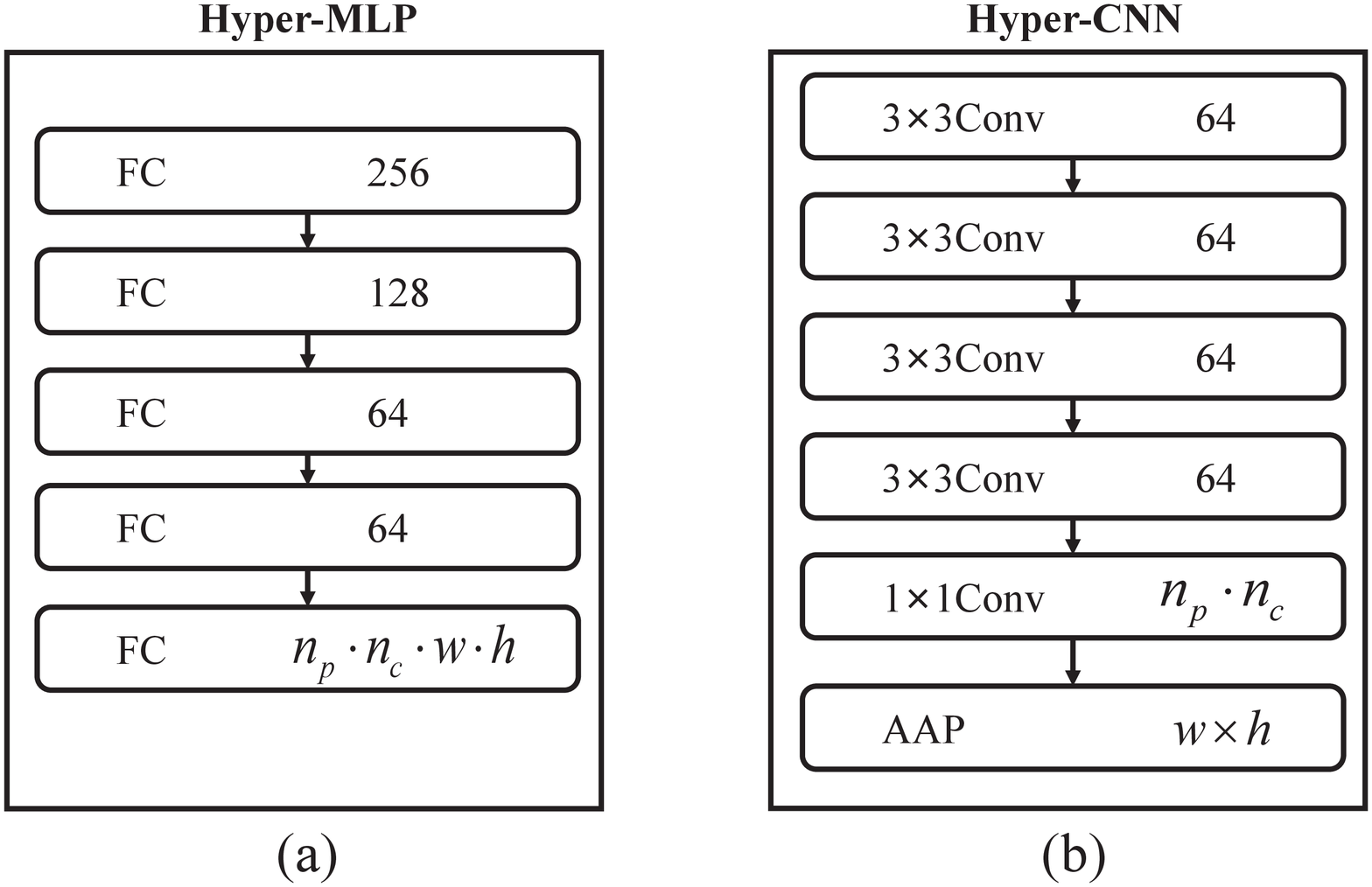}
  \caption{
    MLP or CNN based hypernetwork structure for predicting priors. Where FC is the fully connected layer, Conv is the convolutional layer, and AAP is the adaptive average pooling layer.}
  \label{fig_hypernetworks}
\end{figure}

In fact, human knowledge is not innate and constant, but evolves continuously by summarizing the experience of past tasks. Inspired by this, we borrow the idea of meta-learning \cite{hospedales2021meta} and optimize the MSConv with the given priors as variables in the outer loop, forming a bi-level optimization problem:
\begin{equation}
  \begin{aligned}
    & \omega^*=\arg \min _\omega \sum_{i=1}^M \mathcal{L}^{\text {outer }}\left(\mathcal{D}_{\text {val }}^i ; \theta^{*(i)}(\omega), \omega\right) \\
    & \text { s.t. } \quad \theta^{*(i)}(\omega)=\arg \min _\theta \mathcal{L}^{\text {inner }}\left(\mathcal{D}_{\text {train }}^i ; \theta, \omega\right),
    \end{aligned}
\end{equation}
where $\omega$ represents priors, $M$ is the number of tasks, $\mathcal{L}^{\text{inner}}$ and $\mathcal{L}^{\text{outer}}$ are the objective functions of the inner and outer optimization problems, respectively, $\mathcal{D}_{\text{train} }^i$ and $\mathcal{D}_{\text{val}}^i$ are the training and validation sets for the $i$th task, respectively, and $\theta^{*(i)}(\omega)$ is the optimization parameter for the $i$th task under the guidance of $\omega$. In this bi-level optimization problem, the inner layer is the ordinary image classification problem and the outer layer is the optimization of the knowledge. We constructed an MLP-based and a CNN-based hypernetworks for the outer layer optimization problem, whose structural parameters are shown in Fig \ref{fig_hypernetworks}. They input training samples from the task and output $n_p$ image blocks of size $h\times w$ and number of channels $n_c$ as $\omega_i$, i.e., $\omega_i=\mathcal{H}(\mathcal{D}_i)$. The outer objective function is set to the inner objective value on the validation set.

Few-shot image classification is an image recognition task with very few training samples (generally only 1 to 20 training samples per class), which tests the adaptability of the model to the lack of data volume and its ability to generalize across tasks. In this experiment, we design a semantic recognition model based on an evolvable knowledge base for a small-sample image classification task, and verify the key role of knowledge in achieving a strong interpretable recognition with low sample requirement and high generalization ability. We conducted experiments using the commonly used benchmarks Omniglot \cite{lake2015human} and mini-imagenet \cite{vinyals2016matching}. Omniglot is a handwritten character dataset containing 1623 classes, where each class contains 20 $28\times28$ grayscale images. We use 1200 of these classes for meta-training and 423 classes for meta-testing. Mini-imagenet is a natural image dataset containing 100 classes, where each class contains 600 $84\times84$ RGB images. We construct $T$ task sets for meta-training and meta-testing respectively, where each task set contains 5 random categories with 20 image samples per category. Among them, 5 images from each category are used as training set, 5 as validation set, and 10 as test set. The meta-training has a meta-epoch of 3 and the task training has an epoch of 30. Both the hypernet and MSRM use the Adam optimizer \cite{kingma2014adam} with a learning rate of $1e^{-4}$. We use the average test accuracy on task sets for meta-testing as the performance evaluation criterion.

In order to verify the effectiveness of knowledge evolution based on the hypernetwork, we compared the evolvable priors predicted by the hypernetwork with the fixed priors based on MSRM, and the accuracy is shown in Table \ref{tab_compare_to_fix_priors}. It can be seen that the randomly generated priors have the worst performance among all fixed priors. Since the hypernet is randomly initialized, its initial predicted priors are close to the randomly generated priors, and the performance of the priors predicted by the hypernet after parameter optimization is better than all the fixed priors. The results show that the evolvable knowledge representation can be achieved based on the hypernetwork, and with a small number of samples trained, its performance can exceed that obtained based on manual design or statistical analysis of knowledge.

\begin{table}[]
	  \centering  
 \caption{The accuracy performance comparison between fixed priors and evolvable priors.}
  \begin{tabular}{|l|l|l|}
  \hline
  Prior source/Dataset & Omniglot & Mini-imagenet \\ \hline
  Random               & 0.312        & 0.278             \\ \hline
  Handcrafting         & 0.592        & 0.542             \\ \hline
  Gauss                & 0.546        & 0.536             \\ \hline
  Deconvolution        & 0.834        & 0.736             \\ \hline
  PCA                  & 0.868        & 0.768             \\ \hline
  \textbf{Hyper-FC}    & 0.894        & 0.813  \\ \hline
  \textbf{Hyper-CNN}   & \textbf{0.907}        & \textbf{0.842}  \\ \hline
  \end{tabular}
  \label{tab_compare_to_fix_priors}
\end{table}

We modeled the process by which the intelligence evolves its knowledge to help in the solution of new tasks as its historical experience increases by studying the relationship between the amount of training tasks and performance. The results are shown in Fig. 7. It can be seen that the accuracy rate increases as the amount of tasks increases. This indicates that the intelligence evolves knowledge based on historical experience to benefit the solution of new tasks.
\begin{figure}[htbp]
  \centering	
  \includegraphics{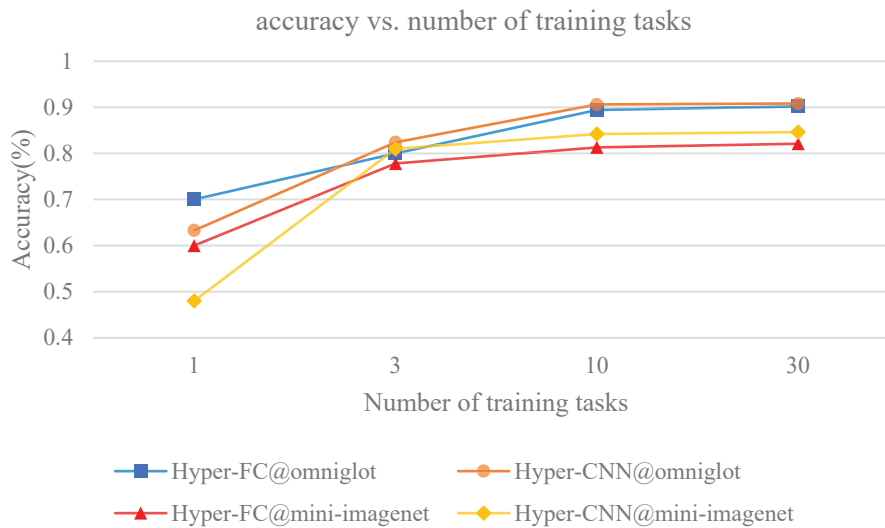}
  \caption{
    Accuracy vs. number of training tasks.}
  \label{fig_acc_tasks}
\end{figure}

		\section{Conclusions}

In this paper,   we proposed
the definitions on message, information, semantics, knowledge and meaning and   their mathematical characterization for
 content-oriented processing.
   Specifically,
 we divide the messages into real-scene messages and virtual-scene messages based on the sources of the messages, where
real-scene message is the signal obtained by sensors detecting objective things, while virtual-scene message is the signal expressed by human beings in language, text, and graphics.
 Then, we defined  meaning as the result of the brain's understanding of the spatio-temporal state of objective
things, and
information is the image of the part of the message that the message receiver does
not know in advance and can decode and understand its meaning.
 Semantics is the label or naming of the characteristic function of the virtual-scene message to the characteristic function of the real-scene message, and knowledge  is the law of the spatio-temporal state of things discovered or created by human beings, which can be expressed by semantic symbols.
  We presented semantic decomposition and composition. We model the semantic decomposition as a bi-level optimization problem based on the principle of complexity and error minimization. By iteratively solving the semantic decomposition problem, a human-like semantic hierarchy can be constructed. By specifying the coefficients of the semantic bases, new semantics can be created and new knowledge can be discovered through environmental verification. Semantic decomposition and composition may provide new ideas and technical traction for achieving human-like intelligence.

			\bibliographystyle{IEEE-unsorted}
			\bibliographystyle{IEEEtran}
\bibliography{reference}

\end{document}